\documentclass[prb,nobibnotes,floatfix,superscriptaddress,twocolumn]{revtex4-2}
\usepackage{amsfonts,amsmath,graphicx,epsfig,latexsym}
\usepackage{subfigure}
\usepackage{graphicx}
\usepackage{color}  
\usepackage{natbib}
\usepackage{multirow}
\usepackage{latexsym,amssymb}
\usepackage{amsmath}
\usepackage{mhchem}
\usepackage{soul}
\usepackage{ulem}
\usepackage{array}
\usepackage{siunitx}
\newcolumntype{d}[1]{D{.}{.}{#1}}
\usepackage[linktocpage,bookmarksopen,bookmarksnumbered]{hyperref}
\usepackage[dvipsnames]{xcolor}

\usepackage{comment}
\usepackage{braket}
\usepackage[utf8]{inputenc}
\DeclareUnicodeCharacter{2032}{'}
\usepackage[utf8]{inputenc}

\newcommand{\dw}{$D(\omega)$}
\newcommand{\rdw}{$\mathfrak R [D(\omega)]$}
\newcommand{\idw}{$\mathfrak I [D(\omega)]$}
\newcommand{\et}{$\eta$}

\usepackage{multirow}

\begin{document}




\title{Accelerated prediction of dielectric functions in solar cell materials with graph neural networks}

\author{Caden Ginter}

\affiliation
{Department of Physics and Astronomy, West Virginia University, Morgantown, WV 26506, USA}

\author{Kamal Choudhary}

\affiliation
{Department of Materials Science and Engineering, Johns Hopkins University, Baltimore, MD 21218, USA}
\affiliation
{Materials Measurement Laboratory, National Institute of Standards and Technology, Gaithersburg, MD 20899, USA}

\author{Subhasish Mandal}
\email{Contact author: subhasish.mandal@mail.wvu.edu}
\affiliation
{Department of Physics and Astronomy, West Virginia University, Morgantown, WV 26506, USA}

\def\dvc#1{\textcolor{red}{[DV: #1]}}
\def\Red#1{\textcolor{red}{#1}}
\def\Blue#1{\textcolor{blue}{#1}}
\def\Magenta#1{\textcolor{magenta}{#1}}

\def\sm#1{\textcolor{cyan}{[SM: #1]}}
\def\smc#1{\textcolor{Brown}{#1}}

\begin{abstract}
{\footnotesize

We present an atomistic line graph neural network (ALIGNN) model for predicting dielectric functions directly from crystal structures. Trained on $\sim$7000 dielectric functions from the JARVIS-DFT database computed with a meta-GGA exchange-correlation functional, the model accurately reproduces spectral features, including peak intensities and overall line shapes, while enabling efficient high-throughput screening. Applied to the recently developed Alexandria materials database, containing over four hundred thousand insulating materials, we uncover a clear elemental trend, with vanadium emerging as a strong indicator of materials with high-spectroscopic limited maximum efficiency (SLME). In particular, vanadium-based perovskite materials show a substantially higher fraction of high-SLME compounds compared to the database average, underscoring their promise for optoelectronic applications.

}
\end{abstract}


\maketitle

\newpage

\section{Introduction}

 The dielectric function is a fundamental property of materials, governing their optical response, electronic excitations, and screening behavior. Accurate knowledge of dielectric function is particularly critical for the design and optimization of optoelectronic devices, photovoltaic and solar cell materials~\cite{PIC-nat,ppv0,perv-ppv1,perv-ppv2,perv-ppv3}. These are one of the most promising avenues for sustainable energy generation and are central to the development of alternative energy technologies across both commercial applications and emerging industries~\cite{Ramayanti_Satellite_Solar2022,Abdelhamid_Car_Solar2014}. 
 The search for improved solar cell materials remains a vibrant and active area within materials science~\cite{nn-ppv1,ppv-research1,ppv-research2,ppv-research3}. Experimentally, the dielectric function can be determined using techniques such as optical reflectivity, spectroscopic ellipsometry, and electron energy-loss spectroscopy. While these approaches provide reliable measurements, they often require highly specialized facilities and high-quality single crystals, which limit their suitability for large-scale, high-throughput evaluations~\cite{die-exp1,die-exp2}.

First-principles methods based on density functional theory (DFT) have enabled computational evaluation of electronic structure, leading to the establishment of large databases (e.g, Materials Project~\cite{mp}, AFLOW~\cite{aflow}, OQMD~\cite{oqmd}, and JARVIS~\cite{Choudhary_Jarvis2020}) encompassing thousands of semiconducting and insulating compounds. However, conventional DFT functionals are known to systematically underestimate electronic band gaps, resulting in appreciable errors in the predicted dielectric response. Beyond-DFT methods, such as the GW approximation combined with the Bethe-Salpeter equation (GW-BSE)~\cite{hedin_new_1965,hybertsen_electron_1986,BGW,SM-GW,PyGW} yield significantly improved accuracy but remain computationally prohibitive for materials with large or complex unit cells due to the necessity of sampling a vast number of unoccupied states~\cite{gw-htp1,gw-database1}. Hybrid functionals~\cite{becke_new_1993,doi:10.1063/1.1564060,PhysRevB.83.235118,SM-mtu1,SM-mtu2} face similar challenges, which constrain their use in building high-fidelity, large-scale dielectric function databases~\cite{hybrid-db}. An emerging alternative is the application of machine learning models trained on curated materials datasets. When developed in conjunction with more accurate exchange-correlation functionals than conventional generalized gradient approximations (GGAs), such models can provide rapid and accurate predictions at a fraction of the computational cost.

Within this context, crystal graph neural networks (CGNNs)~\cite{Xie_CGCNN2018, Zhan2023_CGCNN_perovskites, schutt_schnet2018, chen_m3gnet2022,  Kaundinya_electron_dos2022, Gurunathan_Phonon2023, pati-ANN, Choudhary_ALIGNN2021},  have recently emerged as powerful architectures for learning materials properties directly from atomic structures. By representing atoms as graph nodes and interatomic bonds as edges, CGNNs provide a natural non-Euclidean representation of crystalline solids, enabling accurate prediction of properties such as formation energy, band gap, and elastic constants. Nonetheless, the standard nearest-neighbor connectivity employed in CGNNs often yields an incomplete description of the local chemical environment. To address this limitation, the atomistic line graph neural network (ALIGNN) was introduced as an extension of CGNNs~\cite{Choudhary_ALIGNN2021}. ALIGNN incorporates bond-angle information by constructing a line graph over the original crystal graph: nodes in the line graph correspond to bonds, while edges represent pairs of bonds, thus encoding bond-angle cosines as features. By jointly incorporating both bond connectivity and bond-angle descriptors, the ALIGNN achieves superior predictive accuracy across a broad spectrum of materials properties. These include formation energy, band gap~\cite{Choudhary_ALIGNN2021}, electronic density of states (DOS)~\cite{Kaundinya_electron_dos2022}, phonon DOS and DOS-derived thermodynamic quantities such as heat capacity, vibrational entropy, and isotopic phonon-scattering rates~\cite{Gurunathan_Phonon2023}, as well as properties of two-dimensional van der Waals magnets~\cite{choudhary_magprops_vdW_magnets_GNN} and magnetic moments. In comparative benchmarks, ALIGNN consistently outperforms earlier crystal graph neural network (CGNN) variants-including CGCNN~\cite{Xie_CGCNN2018}, SchNet\cite{schutt_schnet2018}, and M3GNet\cite{chen_m3gnet2022}-as well as descriptor-based approaches that rely on radial or angular distribution functions. 

Despite these advances, large-scale datasets of dielectric functions remain scarce, particularly those obtained with advanced exchange-correlation functionals. Among available approaches, meta-GGA functionals~\cite{mbj,jctc-mbj,mbj-jp,TB-mbj-2,TB-mbj-3,r2-scan,bansil-ph} offer a compelling balance as they provide accuracy for optical properties, including band gaps, that approaches the fidelity of GW calculations, while retaining computational costs comparable to standard GGA~\cite{Metal_paper}. This balance makes meta-GGA a practical and scalable foundation for high-throughput prediction of dielectric and optical properties directly from crystal structures.

In this work, we develop an ALIGNN-based framework specifically designed to predict the dielectric function of insulating materials. The model is trained on the JARVIS-TB-mBJ dataset, which provides electronic-structure calculations using the Tran-Blaha modified Becke-Johnson (TB-mBJ) potential. After establishing and validating this framework, we demonstrate its scalability and practical utility through application to the recently released Alexandria database~\cite{Schmidt_Alexandria2024}, which encompasses more than one million compounds and stands as one of the largest materials databases available today. As a case study, we illustrate how our model can facilitate materials discovery by identifying chemical classes and structural motifs correlated with the spectroscopic limited maximum efficiency (SLME). Notably, approximately 23\% of perovskites in the database exhibit high SLME ($\eta > 25\%$), compared to only 16\% across the entire dataset. Moreover, we identify that the vanadium-based perovskites are particularly promising candidates, showing a higher likelihood of achieving high SLME than perovskites based on other transition metals. Thus, by establishing ALIGNN as a scalable and generalizable framework for predicting complex functional properties, this work aims to extend the role of machine learning in materials science beyond conventional electronic structure descriptors toward predictive modeling of functional optical responses critical for next-generation optoelectronic applications.

\section{Methods}
\begin{figure}[!htb]
        \includegraphics[width=1.0\linewidth]{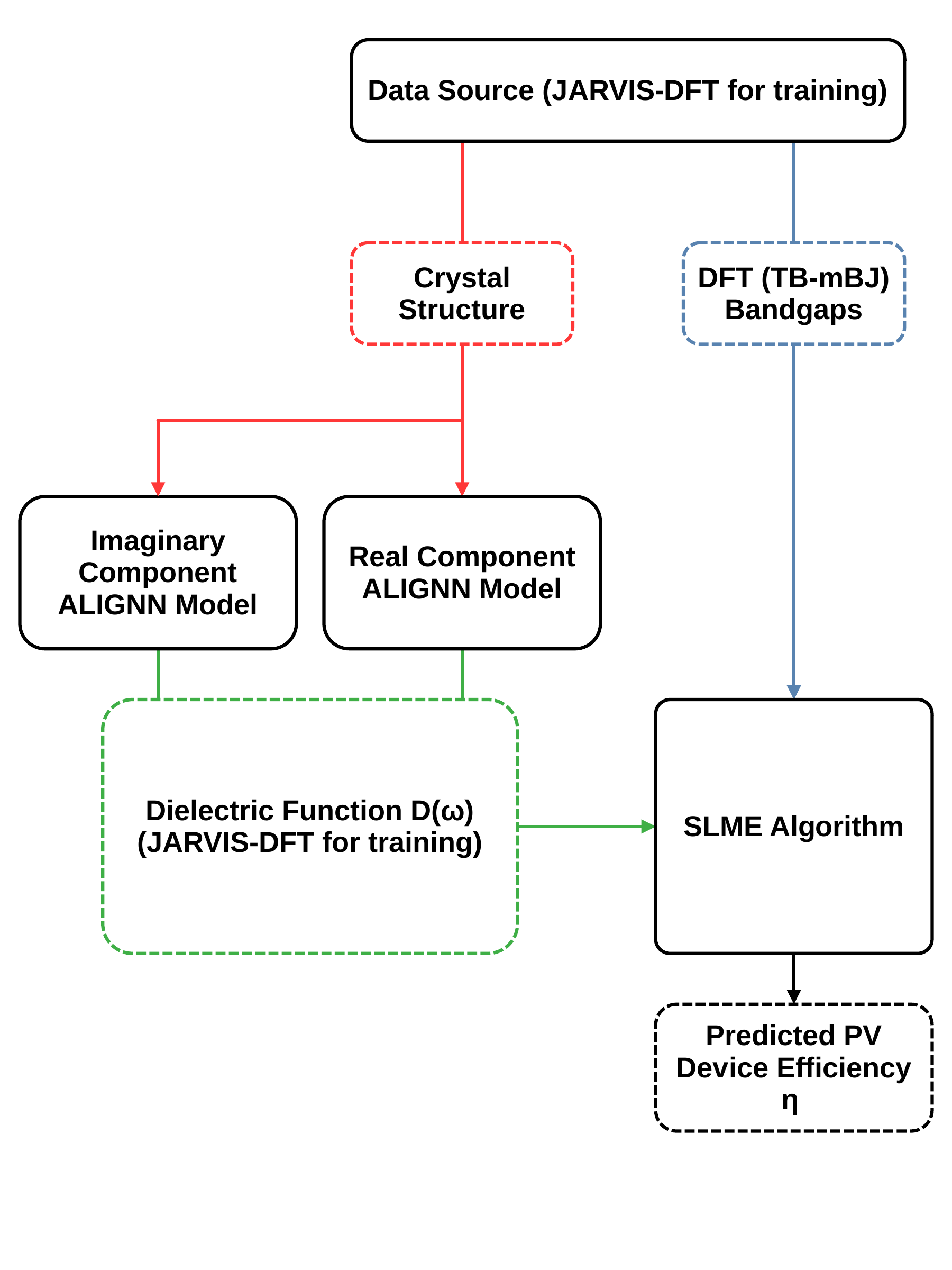}
        \caption{Flow chart demonstrating the proposed technique, where software and data transfer components are represented in solid and dashed boxes respectively.}
        \label{fig:diagram}
    \end{figure}

In this section, we outline the construction of the machine-learning models, followed by a description of the data acquisition and pruning procedures. Next, we describe the training procedure and the computation of SLME. We then detail the evaluation methods used to assess model performance, both with respect to the training data and in terms of the SLME metric. The overall workflow for model training and application is summarized schematically in Fig.~\ref{fig:diagram}. The crystal structures and bandgap data are obtained from the JARVIS-DFT database using the TB-mBJ functional. During training, the models learn to predict the dielectric function directly from crystal structure information, using the TB-mBJ-calculated dielectric functions as targets. Two separate ALIGNN models are employed-one for the real and one for the imaginary component of the dielectric function, \dw{}, each optimized independently for its respective output. Once trained, the workflow can be applied to any material with a known crystal structure and bandgap. To estimate the theoretical solar efficiency limit, \et{}, the crystal structure is first passed through the trained models to generate its \dw{}. This predicted dielectric function, together with the direct and indirect bandgaps, is then used within the SLME framework to compute \et{}. The entire process is automated and readily scalable to large materials datasets with precomputed bandgaps, such as the Alexandria database.

    \subsection{Machine Learning Models}
    Our models for predicting the dielectric function \dw{} of a given material structure are built using the Atomistic Line Graph Neural Network (ALIGNN) framework ~\cite{Choudhary_ALIGNN2021}. This framework allows a neural network to operate on a non-Euclidean graph representing the crystal. The key advantage of this framework, and the motivation for adopting it here, is its explicit encoding of bond-angle information through a secondary, line-graph representation. A detailed description of the ALIGNN architecture is provided in Ref.~\cite{Choudhary_ALIGNN2021}. An additional challenge arises from the fact that the dielectric function is complex valued~\cite{GriffithsElectrodynamics2014,ashcroft1976solid}. Several strategies can be envisioned to address this, including encoding the real and imaginary components within a single model or using analytical continuation to derive one component from the other~\cite{nn-ppv1}. In the present study, we adopt a simpler and flexible approach in which two independent models are trained: one for the real part, \rdw{}, and one for the imaginary part, \idw{}. The full dielectric function is then reconstructed by combining predictions from both models. The input to each model is the same structure, with conversion to graph representation handled by the ALIGNN framework. Each model outputs 300 discrete samples of its component of \dw{}, sampled at a regular interval of 0.05 eV, spanning from 0 eV to 15 eV.


    \subsection{Data Acquisition}

  Our training and test datasets are obtained from the JARVIS-DFT repository, which provides DFT calculations for a wide range of materials with the TB-mBJ functional. Specifically, we use data for 6918 bulk crystalline compounds, each associated with its computed frequency-dependent dielectric function, \dw{}, and electronic band gap, both evaluated using the TB-mBJ functional. In addition, we retrieve the reference SLME values from the same database for use in downstream validation. 

  During the initial stages of model development, we found that several entries in the raw dataset contained numerical inconsistencies or nonphysical features in the reported dielectric functions. To address this issue, we implement a filtering procedure that excluded any material meeting at least one of the following criteria: (i) the band gap should be at least 0.15 eV to ensure the system to be insulating; (ii) one or more dielectric function values are reported as \verb|NaN|;  (iii) all values of the dielectric function are identically zero; (iv) the imaginary component of the dielectric function exceeded 5.0 within the band gap region; or (v) the crystal structure could not be reconstructed from the database entry. After applying this filtering protocol, we obtain a curated dataset of 6918 insulating crystal structures with corresponding TB-mBJ dielectric functions, band gaps, and SLME values. This dataset forms the basis for training and evaluating the ALIGNN models described in this work.

    \subsection{Model Training and Usage}
    
   We train two ALIGNN models, one for the real part (\rdw{}) and one for the imaginary part (\idw{}) of the dielectric function. Both models are implemented using the official ALIGNN framework~\cite{Choudhary_ALIGNN2021}, with hyperparameters adapted from prior benchmarks. Crystal graph neural network (CGCNN) features~\cite{Xie_CGCNN2018} are used to represent atomic environments, and the architecture employed two ALIGNN convolution layers and two graph convolution layers. Each model produced 300 output features to match the dimensionality of the dielectric function dataset. The predicted dielectric functions are subsequently used to evaluate the SLME, a theoretical upper bound on solar cell performance that generalizes the Shockley-Queisser (SQ) limit~\cite{Shockley_SQ1961}. The SLME calculations are performed using the JARVIS-tools implementation~\cite{Choudhary_Jarvis2020,choudhary2025jarvis}, which requires as input the complex dielectric function as well as the direct and indirect band gaps of the material as described below.

\subsection{Computations of SLME}
    The Spectroscopic Limited Maximum Efficiency (SLME) measures the theoretical limit to solar efficiency of a material. This boundary is computed as follows\cite{Choudhary_SolarQuantum2019, Choudhary_Jarvis2020}, where the dielectric function is defined as:

    \begin{equation}
        D_{\alpha \beta}^{(1)}(E) = 1 + \frac2\pi P\int_0^\infty \frac{\epsilon_{\alpha \beta}^{(2)} (E') E'}{(E')^2 - E^2 + i \beta} \mathrm{d} E'
    \end{equation}
    
    \begin{equation}
        \begin{split}
            D_{\alpha \beta}^{(2)} (E) =& \frac{4 \pi^2 e^2}{\Omega^2} \lim_{q \rightarrow 0} \frac1{q^2} \sum_{c,v,\vec k} 2 w_{\vec k} \delta (\xi_{c \vec k} - \xi_{v \vec k} - E) \\ &\braket{\Psi_{c \vec k + \vec e_\alpha q} | \Psi_{v \vec k}} \braket{\Psi_{v \vec k}| \Psi_{c \vec k + \vec e_\beta q}}^{*}
        \end{split}
    \end{equation} \\

    Where $e$ is the electron charge, $\Omega$ is the cell volume, $E$ is the energy, $\beta$ is the broadening, Fermi-weights are $w_{\vec k}$, and $e_\alpha$ are the three Cartesian unit vectors. $\ket{\Psi_{n \vec k}}$ represents the periodic part of the wavefunction for band $n$ and k-point $k$. The incident wave's Bloch vector is $q$. $c$ and $\xi_{c \vec k}$ are the conduction band and its eigenvalue, respectively, for a given k-point, while $v$ and $\xi_{v \vec k}$ are defined similarly for the valence band. $D_{\alpha \beta}^{(1)}(E)$ is derived from $D_{\alpha \beta}^{(2)}(E)$ by the Kramers-Kronig transformation. The tensors resulting from these computations are diagonalized to obtain $D^{(1)}$ and $D^{(2)}$, which are used to compute the absorption coefficient as\cite{Choudhary_SolarQuantum2019},

    \begin{equation}
        \alpha (E) = \frac{2 E}{\hbar c} \sqrt{\frac{\sqrt{[D^{(1)}(E)]^2 + [D^{(2)}(E)]^2} - [D^{(1)}(E)]}{2}}
    \end{equation}

    The SLME (\et{}) is then defined as

    \begin{equation}
        \eta = \frac{P_\text{max}}{P_\text{in}}
    \end{equation}

    where $P_\text{max}$ is defined as the maximum power output of the device and $P_\text{in}$ is the power of the incident solar radiation on the surface of the device. Thus, \et{} is the maximum proportion of incident solar power that the device can capture and supply as electrical power. The SLME calculations are carried out using the JARVIS-Tools infrastructure~\cite{Choudhary_Jarvis2020}, which incorporates the methodology developed in Ref.~\cite{Yu_SLME2012}, enabling an efficient and materials-specific evaluation of photovoltaic performance.

 \subsection{Model Evaluation}

We employ two complementary metrics to evaluate model performance. The first is a normalized error metric defined as the ratio between the mean absolute error (MAE) of the prediction and the mean absolute deviation (MAD) of the reference data:  

\begin{equation}
\text{MAE:MAD} = 
\frac{\frac{1}{N} \sum_{i=1}^{N} |y_i - \hat y_i|}
{\frac{1}{N} \sum_{i=1}^{N} |y_i - \mu|}, 
\quad \mu = \frac{1}{N} \sum_{i=1}^{N} y_i ,
\end{equation}

where $y_i$ are the discrete samples of the reference function, $\hat y_i$ are the corresponding predicted values, and $N=300$ is the number of sampled energy points. Lower MAE:MAD values indicate improved model performance, with the normalization ensuring robustness across functions of different spreads.  

To further quantify the alignment of spectral features, we adopt the Cosine Similarity of Derivatives (CSD). We approximate smoothed finite-difference derivatives of the reference and predicted functions with a frequency grid size of 0.5 eV as

\begin{equation}
y_i' \approx \frac{y_{i+10} - y_i}{0.5 \,\text{eV}}, \quad 
\hat y_i' \approx \frac{\hat y_{i+10} - \hat y_i}{0.5 \,\text{eV}},
\end{equation}

 This grid size is selected so that the CSD metric represents function-fit in the frequency regime of interest. We then compute the cosine similarity of the resulting vectors:  

\begin{equation}
\text{CSD} = 
\frac{\sum_{i=1}^{N-10} y_i' \, \hat y_i'}{
\sqrt{\sum_{i=1}^{N-10} (y_i')^2} \,
\sqrt{\sum_{i=1}^{N-10} (\hat y_i')^2}} .
\end{equation}

CSD values closer to 1 indicate strong agreement in peak and valley alignment, whereas negative values indicate opposite trends in spectral behavior.  

\subsection{Alexandria Database Analysis}

We apply the trained models to the recently released Alexandria database~\cite{Schmidt_Alexandria2024}, which contains approximately 5 million crystalline materials. Out of this dataset, we select the PBE-computed 3D materials. From this set, 438{,}216 materials are selected based on two criteria: a direct band gap greater than 0.15~eV and successful reconstruction of a valid atomic structure. Direct and indirect band gaps are also retrieved for subsequent SLME calculations. To investigate chemical trends, we further identify perovskite structures within this dataset. Perovskites are loosely defined according to the general formulas $\text{A}_n \text{T}_n \text{X}_{3n}$ and $(\text{A}_n \text{T}_n \text{X}_{3n})(\text{B}_n \text{T}_n \text{X}_{3n}) 
= \text{A}_n \text{B}_n \text{T}_{2n} \text{X}_{6n}$, where A and B denote arbitrary cations, T a transition metal, X a chalcogen, and $n \geq 1$. This approach to perovskite definition is based purely on stoichiometry and thus does not consider the actual structural parameters of the material, unlike newer tolerance factor approaches that require ionic charges \cite{turnley_perov_tolerance_factor2024}.  For ambiguous cases where multiple assignments of A, B, and T are possible, structures are counted once in the total perovskite population but included in each relevant subclass. This approach allowed us to partition statistics by transition metal and by chalcogen while avoiding double-counting in the overall totals.

\section{Results \& Discussions}

\subsection{Model Evaluation}
We assess the performance of the trained models using multiple metrics. Each model is evaluated individually based on the MAE:MAD, and CSD measures, followed by an analysis of their combined performance within the SLME framework. All evaluations are carried out on a test set comprising 10\% of the filtered dataset, randomly selected at training time by the ALIGNN framework. The full list of materials in the test set is available through our GitHub repository~\cite{ALIGNN_opto_github}.

    \begin{figure}[!htb]
        \includegraphics[width=240pt, angle=0]{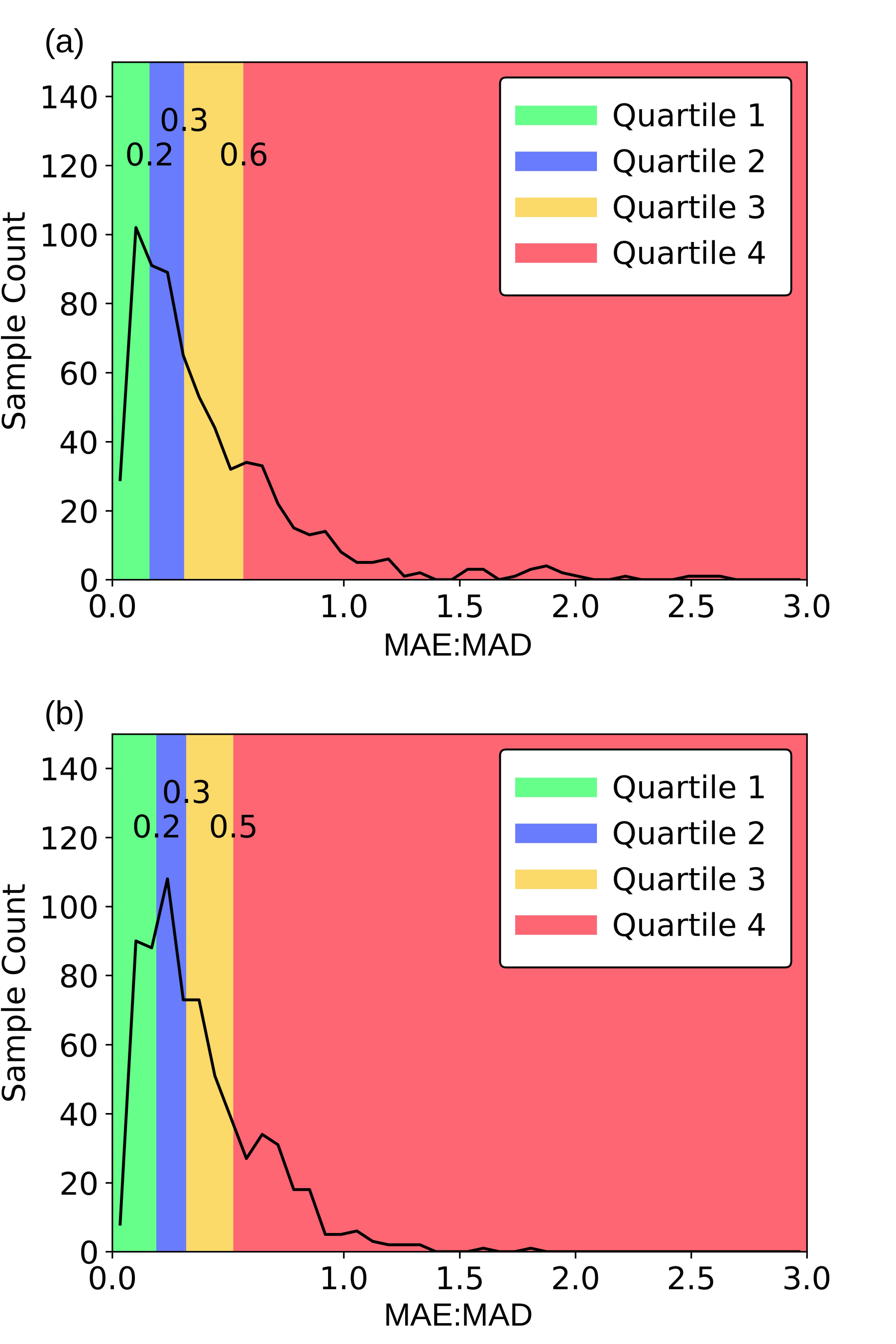}
        \caption{Distribution of MAE:MAD scores across the test set for the proposed model in obtaining (a) imaginary part (\idw{}) \& (b) real part (\rdw{}) of the dielectric function; quartiles are indicated as shaded regions.}
        \label{fig:error_dist}
    \end{figure}

    \begin{figure*}
        \includegraphics[width=440pt, angle=0]{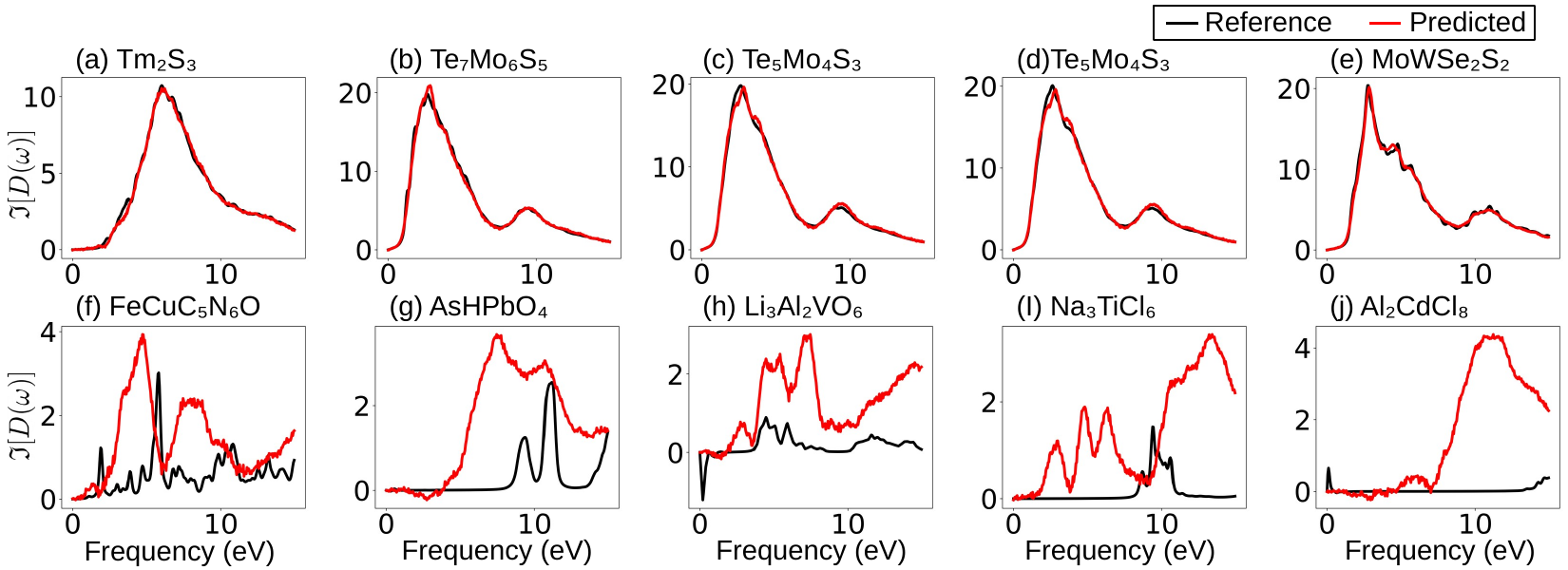}
        \caption{5 best (top) and 5 worst (bottom) examples from the performance of \idw{} model. Red and black represent predicted and reference data respectively. }
     
        \label{fig:IMAG_bestworst}
    \end{figure*}

    \begin{figure*}
        \includegraphics[width=440pt, angle=0]{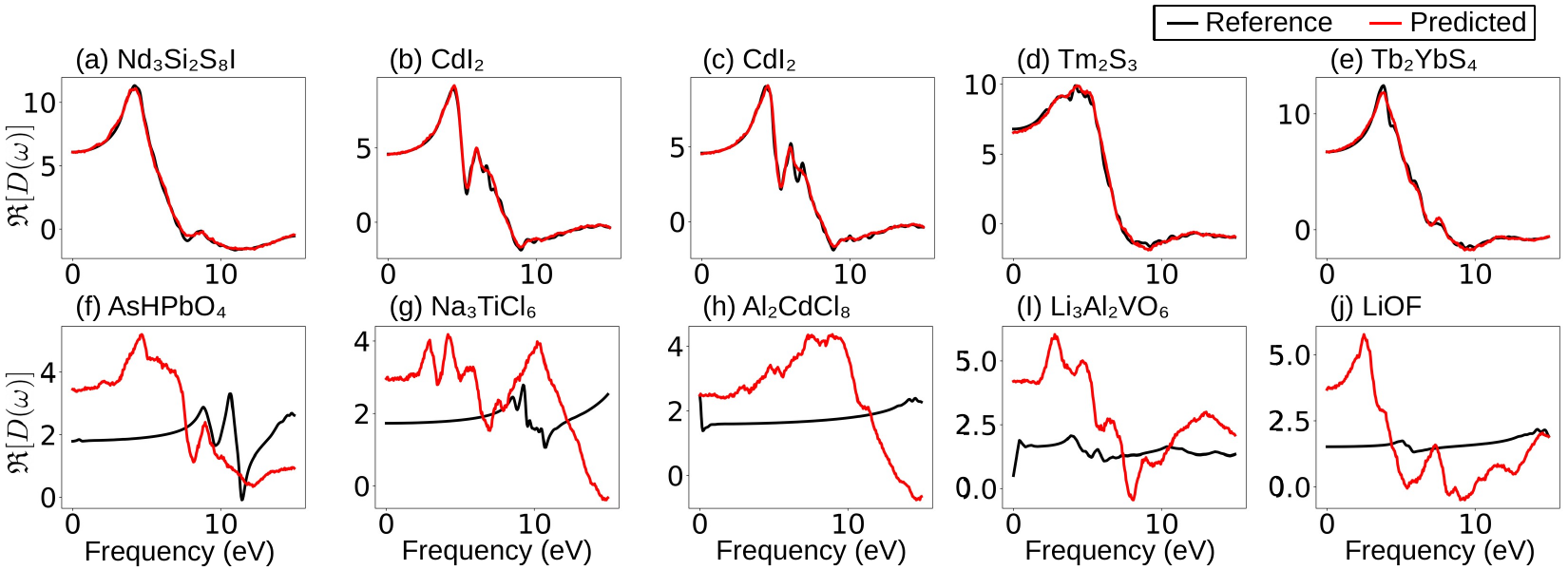}
        \caption{ 5 best (top) and 5 worst (bottom) examples from the performance of \rdw{} model. Red and black represent predicted and reference data respectively.}

        \label{fig:REAL_bestworst}
    \end{figure*}

Figure~\ref{fig:error_dist} presents the distribution of MAE:MAD values across the test dataset for both the real and imaginary component models. For each test sample, the corresponding model prediction is compared with reference database values using both the MAE:MAD and CSD metrics. The MAE:MAD results are shown as line plots with sample counts binned by score, while shaded regions denote quartiles of the distribution. In both cases, the distributions are right-skewed, reflecting a clustering of high-quality, low-error predictions. Although the two models perform comparably, notable differences are observed. The \idw{} model [Fig.~\ref{fig:error_dist}(a)] exhibits a broader distribution with a lower peak and a greater number of outliers in the 1.5-2.0 range compared to the \rdw{} model [Fig.~\ref{fig:error_dist}(b)].

We also report 95\% confidence intervals for each evaluation metric. For the \idw{} model, the population mean of MAE:MAD lies between 0.36 and 0.55 with 95\% confidence, while the corresponding interval for CSD is 0.47-0.53. For the \rdw{} model, the MAE:MAD mean lies between 0.42 and 0.52 at the same confidence level, and the CSD mean is bounded between 0.43 and 0.55.

Figures \ref{fig:IMAG_bestworst} and \ref{fig:REAL_bestworst} show example results from the imaginary and real models, respectively. These figures illustrate the character of the model's performance at both ends of the spectrum. 
In each figure, subpanels (a-e) correspond to the five materials with the lowest (best) MAE:MAD scores for the respective model, while subpanels (f-j) correspond to the five highest (worst) scores. Figure~\ref{fig:IMAG_bestworst} illustrates the performance of the \idw{} model in both regimes. Materials with low error (a-e) exhibit qualitatively similar features in their imaginary dielectric functions, whereas those with high error (f-j) display significantly greater variability. This trend suggests that the \idw{} model performs particularly well for systems resembling compounds such as \ce{Tm2S3}, \ce{Te7Mo6S5}, and others in the top row, but is less reliable for materials with more diverse spectral features.  Figure~\ref{fig:REAL_bestworst} presents analogous results for the \rdw{} model. Unlike the \idw{} case, the \rdw{} model demonstrates greater qualitative variety among well-predicted materials, suggesting reduced bias and improved generality when modeling the real component of the dielectric function.

\begin{figure}[!htb]
        \includegraphics[width=240pt, angle=0]{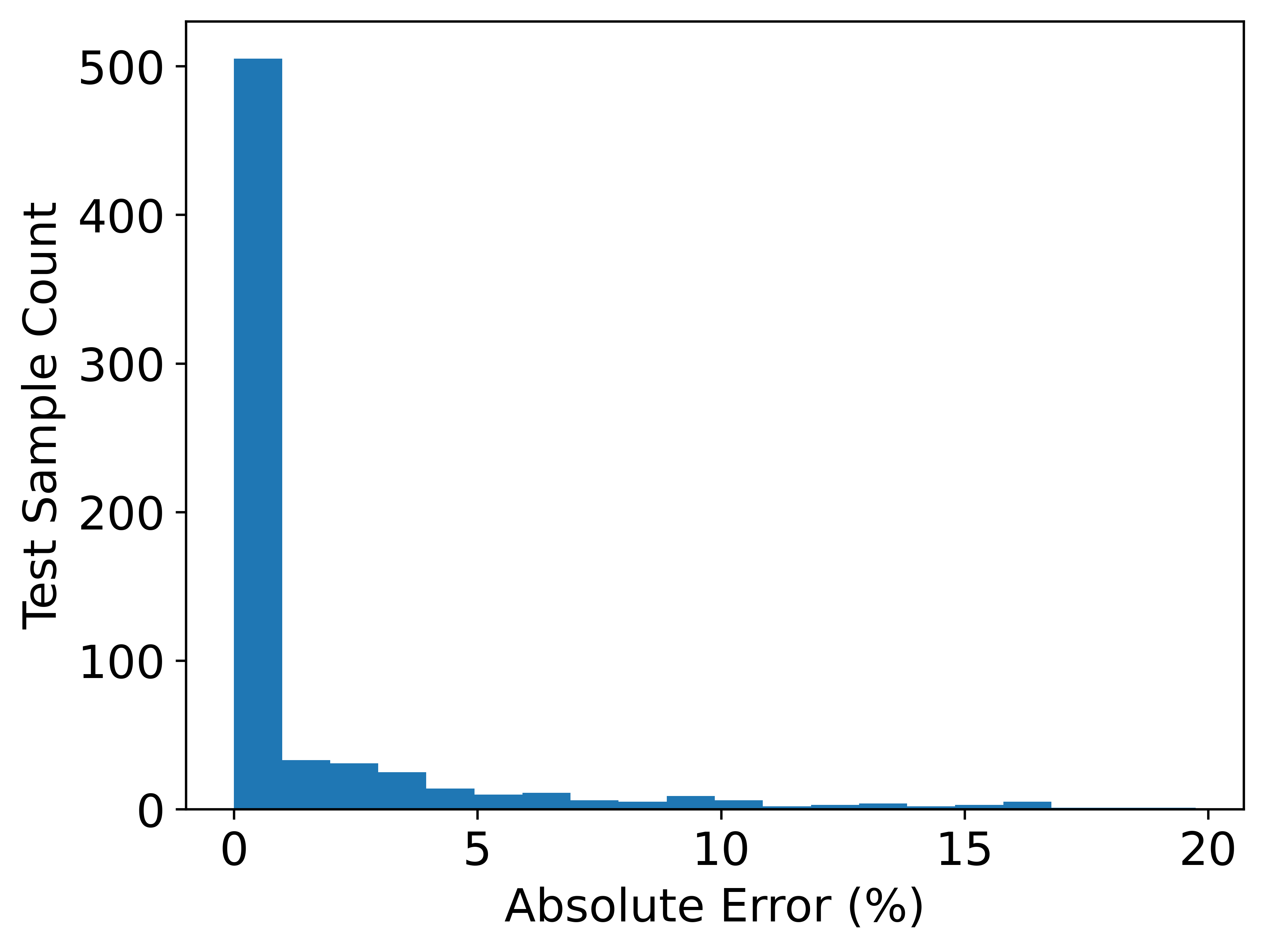}
        \caption{Distribution of the absolute errors in Spectroscopic Limited Maximum Efficiency (SLME) across the test set with MAE of 1.95\%.}
        \label{fig:SLME_test_hist}
    \end{figure}

\begin{figure}[!htb]
        \includegraphics[width=240pt, angle=0]{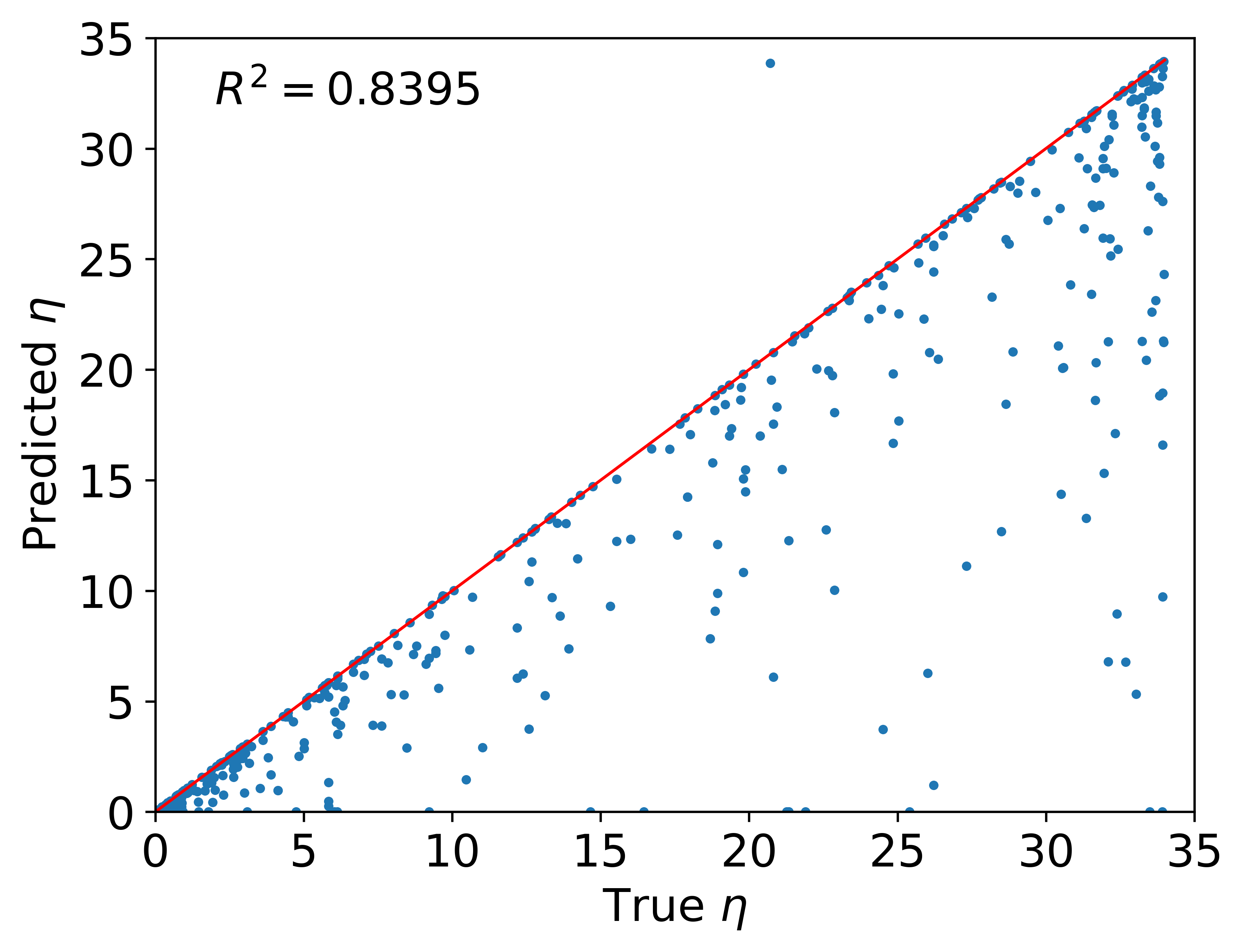}
        \caption{Plot of the predicted \et{} compared with the reference \et{}, yielding $R^2 = 0.8395$. The identity function is plotted on top for reference.}
        \label{fig:r_sq}
    \end{figure}

In addition to analyzing each model separately, we evaluate the models' functionality as a drop-in replacement for traditional DFT and beyond-DFT techniques in computing the SLME of each material. More specifically, for each material, we infer with both models (\idw{} and \rdw{}) to get the full complex-valued \dw{}. This result is then used in place of traditionally-computed \dw{} as input to the JARVIS-tools implementation of SLME \cite{Choudhary_Jarvis2020,Choudhary_SolarQuantum2019}. We use our models' predicted \dw{} with DFT-computed bandgaps with the SLME technique to find the theoretical maximum efficiency $\eta$ of the material. We assess the accuracy of this hybrid technique's $\eta$ using the test data partitioned by the ALIGNN framework. Since the models share the same set of test materials (that were not used in training), we can use that same data to test the machine learning-augmented SLME workflow. These results are summarized in Figure \ref{fig:SLME_test_hist}, which is a histogram of absolute errors across the test set. The mean absolute error (MAE) of this distribution is 1.952, with a 95\% confidence interval placing the true population MAE between 1.60\% and 2.30\%. Thus, on average, the machine-learning-augmented SLME workflow predicts $\eta$ within 2.30\% of the traditionally computed values. This level of accuracy indicates that the coupled models can serve as effective drop-in replacements for DFT or beyond-DFT calculations of \dw{} in the SLME framework. While not exact, the relatively small error demonstrates the practicality of employing the augmented workflow in screening tasks central to novel materials discovery. We further explore the performance of the full SLME workflow in Figure \ref{fig:r_sq}, which illustrates the correlation between the true reference \et{} and the \et{} predicted by the SLME workflow. Besides one outlier, the model is very conservative, under-predicting the efficiency with the  $R^2$ value of 0.8395. This is slightly higher than a recent prediction where graph neural networks were used to train dielectric functions obtained using the GGA dataset from Materials Project~\cite{nn-ppv1}. Additionally, the obtained behavior in Figure \ref{fig:r_sq} is particularly important when the model is employed to pre-screen materials for high-\et{} performance, as it may fail to identify certain true high-\et{} candidates but is unlikely to produce false positives.

    \subsection{Analysis on Alexandria Database}

       \begin{figure*}
            \includegraphics[width=\textwidth, angle=0]{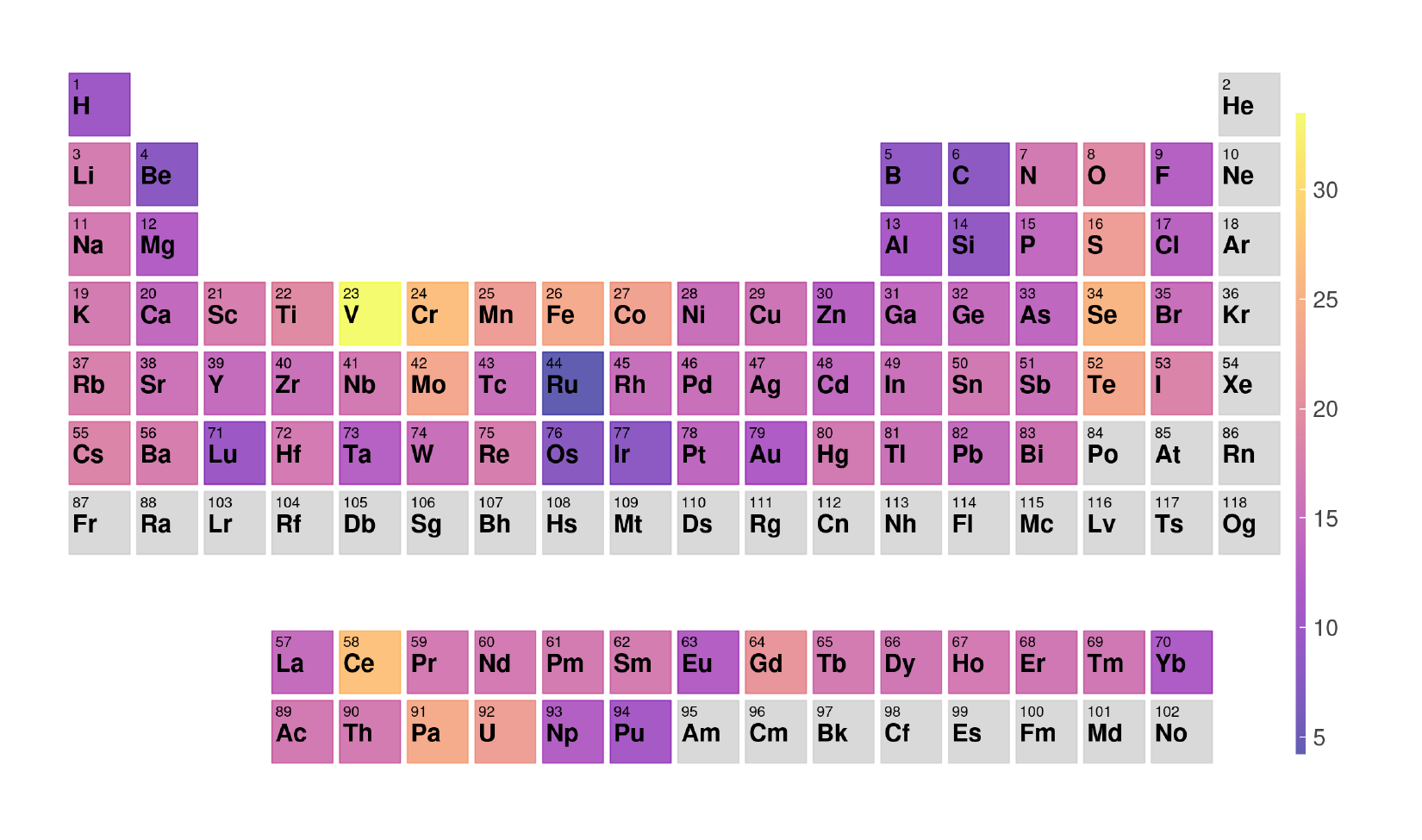}
            \caption{ Intensity map for the percentage of materials containing an element in the periodic table with high SLME ($\eta > 25\%$); yellow and blue represent elements with the largest and the lowest proportion of high SLME materials respectively. Elements shaded in gray are excluded because of low statistical relevance due to small sample size.
        }
            \label{fig:ALEX_periodic}
        \end{figure*}

\begin{figure}[!htb]
            \includegraphics[width=240pt]{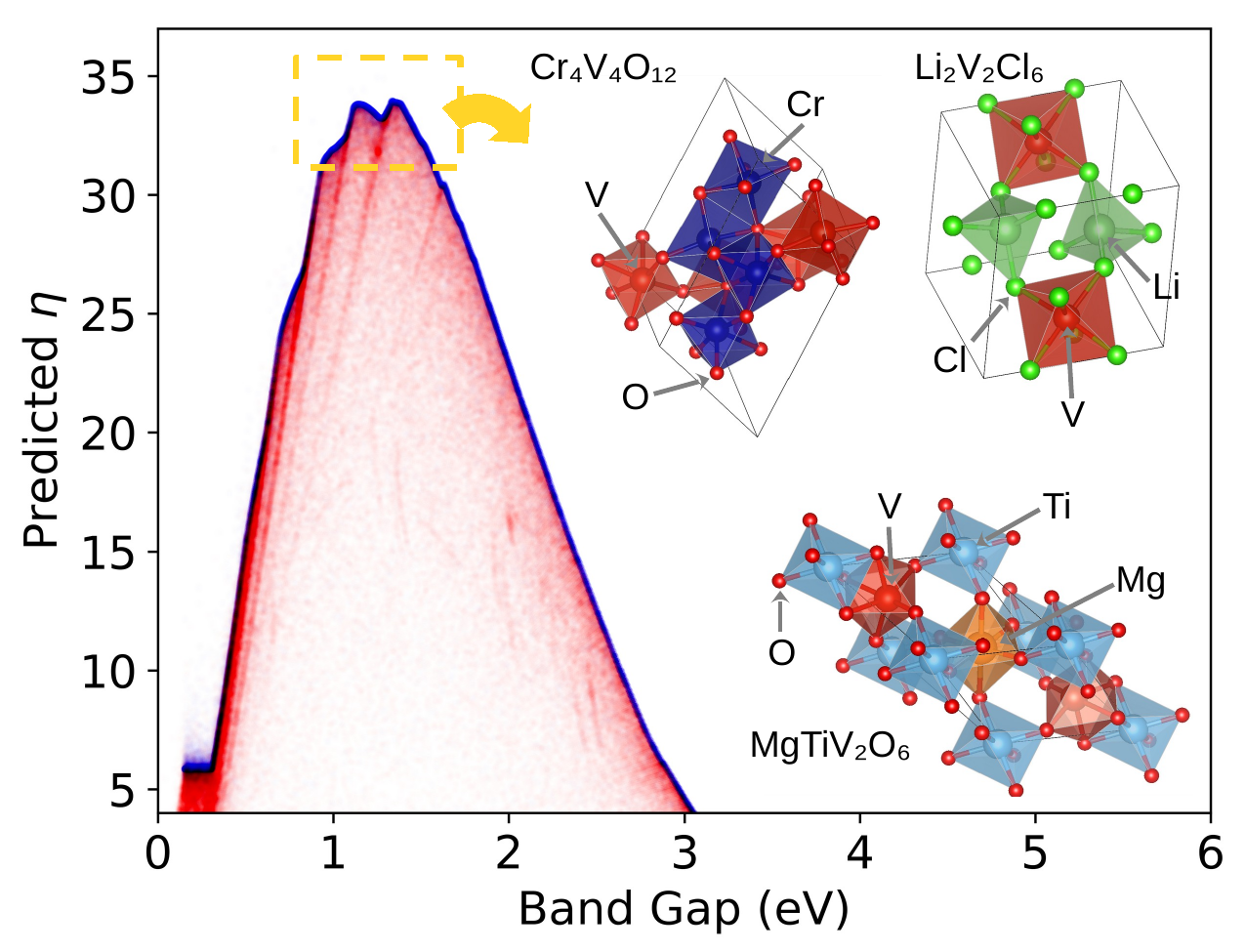}
            \caption{Scatter plot comparing predicted \et{} to DFT band gap for the Alexandria database. Points plotted in blue (red) are for direct (indirect) bandgap materials. The dense portion of the upper boundary of the data roughly follows the traditional Shockley-Quiesser metric curve. The inset shows the crystal structures of a few high-SLME V-based perovskites. }
            \label{fig:eta_vs_bandgap}
        \end{figure}

        We apply our technique to roughly 440,000 insulating materials from the Alexandria database~\cite{Schmidt_Alexandria2024} of 3D crystalline materials. For each material, we predict its dielectric function \dw{} and use this to compute its $\eta$ with the SLME technique. We also classify each material as high or low SLME using a threshold of 25\% efficiency. Materials with $\eta$ greater (lesser) than 25\% are considered to have high (low) SLME efficiency. The top 5 highest-SLME materials from this analysis are: \ce{Pm3SmS6}, \ce{CaZnPF}, \ce{RbTmGeSe4}, \ce{PCa2F}, \ce{CaHo2SnN4}. We first evaluate the density of high-SLME materials containing each element, which is illustrated with the periodic table plot in Figure \ref{fig:ALEX_periodic}, showing the percentage of materials containing each element that are classified as high-SLME. This visualization highlights groups of elements that are more likely to occur in materials with high predicted SLME. In particular, $d$-shell transition metals such as V, Cr, Mn, Fe, and Co, as well as chalcogens O, S, Se, and Te, appear prominently. These high-density regions are consistent with the elemental composition of perovskites, motivating a focused analysis of this material class. We also demonstrate in Figure \ref{fig:eta_vs_bandgap} the relationship between \et{} and the bandgap of materials in the Alexandria database. This figure distinguishes between direct and indirect bandgap materials by color, with blue representing direct bandgap samples and red representing indirect bandgap samples. Samples are considered direct bandgap if $E_g^d - E_g^i < 0.001$, where $E_g^d$ and $E_g^i$ are the direct and indirect bandgap, respectively. These materials are considered Optical Type 1 in the definition of SLME \cite{Yu_SLME2012}, and are thus predicted well by the traditional SQ metric. Indirect bandgap materials span Optical Types 2-4 and require the more advanced methods of SLME to accurately determine their efficiency \et{}. In Figure \ref{fig:eta_vs_bandgap}, these direct bandgap materials (rendered in blue) follow the traditional SQ metric curve, while the indirect bandgap materials (rendered in red) deviate from it significantly. These trends are consistent with the expected behavior discussed in Ref.~\cite{Yu_SLME2012}, which, however, does not explicitly address the low-\et{} regime.

From the Alexandria database, we have identified 10,702 perovskites, corresponding to approximately 2.46\% of the materials analyzed. Among these, 22.9\% exhibit high SLME ($\eta > 25\%$), compared to only 16.3\% across the full set of 435,410 materials. The substantially larger fraction of high-SLME compounds within the perovskite subset indicates a strong correlation between the perovskite structure and enhanced solar efficiency. To give a few examples of higher SLME perovskites, we identify \ce{MgTiV2O6}, \ce{Cr4V4O12}, and \ce{Li2V2Cl6} to have SLME of 33.95\%, 33.89\%, and 33.47\%, respectively. Their crystal structures are shown in the inset of Figure \ref{fig:eta_vs_bandgap}.

   \begin{figure}[!htb]
        \includegraphics[width=240pt]{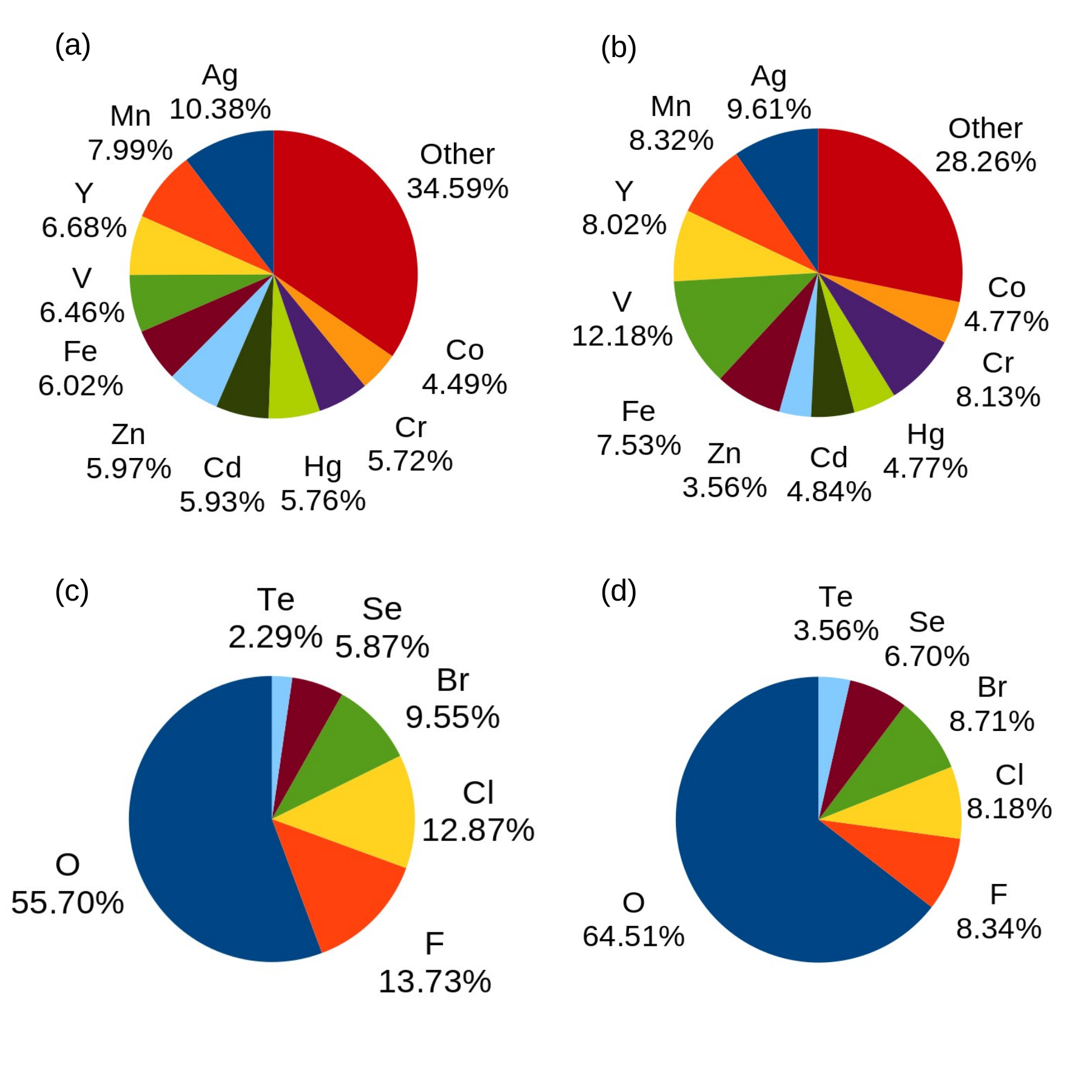}
 
        \caption{
        Pie charts demonstrating the proportion of each type of perovskite in both the entire perovskite collection and the high SLME portion of the collection. (a) proportion of each transition metal in the collection of perovskites, (b) proportion of each transition metal in the collection of high-SLME perovskites, (c) proportion of each chalcogen in the collection of perovskites, (d) proportion of each chalcogen in the collection of high-SLME perovskites
        }
        \label{fig:pies}
    \end{figure}

 We then partition the perovskites based on their type. We partition them in two ways: by their transition metal (Figure \ref{fig:pies} a-b), and by their chalcogen (Figure \ref{fig:pies} c-d). Figure \ref{fig:pies} shows these partitions, each with a chart for all perovskites [Figure \ref{fig:pies} (a, c)] and a chart that counts only high-SLME perovskites [Figure \ref{fig:pies}(b, d)] for comparison. A particularly stark difference exists between the charts partitioned by transition metal, specifically with the element Vanadium (V). Elements V, Cr, and Y have a greater presence in the high-SLME portion of the perovskites set than they do in the full perovskites set, indicating that there may be an additional correlation between a perovskite having these elements as the transition metal and being high-SLME. Figure \ref{fig:ALEX_periodic} agrees with this result, indicating that V has a high density of high-SLME materials in general. The most prominent Vanadium perovskite is \ce{AVO3}, with 511 detected perovskites. Of these, 55.8\% are classified as high-SLME. The results in Figure \ref{fig:pies} also show some interesting behavior in the partitions by chalcogen. Cl, F, and especially O all have significantly higher presence in the high-SLME subset of the perovskites set than they do in the full perovskites set, suggesting a relationship between these chalcogens and high-SLME materials.

\section{Conclusions}

Our results demonstrate that the ALIGNN framework provides an efficient and accurate approach for modeling dielectric functions, enabling high-throughput exploration of large materials databases. Application to the Alexandria database revealed clear elemental trends, with vanadium emerging as a strong indicator of high-SLME materials. Perovskites, in particular, show a markedly higher fraction of high-SLME compounds compared to the database as a whole, underscoring their promise as optoelectronic materials. Taken together, these findings highlight the power of machine-learning augmented workflows to uncover design principles and guide the discovery of next-generation dielectric materials. Additionally, our approach can be extended to transfer learning~\cite{Gupta2021Nov,hoffmann_transfer_learning} for more complex systems, such as heterostructures, layered materials, and to Beyond-DFT databases~\cite{TMO1-SM}. Since most of the high-SLME materials contain transition metals, our work invites high-fidelity high-throughput computation based on dynamical mean field theory as was recently applied to screen altermagnetic materials~\cite{prl-am}.

\section{Acknowledgments } 
This work was supported by the National Science Foundation  (Grants No. NSF OAC-2311558). Computational resources were provided by the WVU Research Computing Dolly Sods HPC cluster, which is funded in part by NSF OAC-2117575 and the Frontera supercomputer at the Texas Advanced Computing Center (TACC) at the University of Texas at Austin, which is supported by National Science Foundation Grant No. OAC-1818253. K.C. thanks the National Institute of Standards and Technology for funding, computational, and data-management resources. This work was performed with funding from the CHIPS Metrology Program, part of CHIPS for America, National Institute of Standards and Technology, U.S. Department of Commerce. Certain commercial equipment, instruments, software, or materials are identified in this paper in order to specify the experimental procedure adequately. Such identifications are not intended to imply recommendation or endorsement by NIST, nor it is intended to imply that the materials or equipment identified are necessarily the best available for the purpose.

\vspace{0.2 in}
\textbf{Data availability}
Our training and testing data is from the JARVIS-DFT database \cite{Choudhary_Jarvis2020}. The collection we applied our model to is the freely available Alexandria \cite{Schmidt_Alexandria2024} database, accessed through JARVIS-tools. The code and data used in this study is available at this
GitHub repository~\cite{ALIGNN_opto_github}.

\section{References} 
\bibliography{CGbib.bib}

\end{document}